\documentclass{ws-procs9x6}

\begin{document}

\title{Isospin in fragment production}

\author{V. BARAN}

\address{Laboratorio Nazionale del Sud, Via S. Sofia 44,
I-95123 Catania, Italy\\ NIPNE-HH and University of Bucharest, Romania\\
E-mail:baran@theor1.theory.nipne.ro}
 
\author{M. COLONNA, M. DI TORO and V.GRECO}

\address{ Laboratorio Nazionale del Sud, Via S. Sofia 44,
I-95123 Catania, Italy\\ and University of Catania\\
E-mail:ditoro@lns.infn.it}  

\maketitle

\abstracts{Based on a general approach to binary systems we show that
in low density region asymmetric nuclear matter (ANM) is unstable
only against isoscalarlike fluctuations. The physical meaning of
the thermodynamical chemical and mechanical instabilities is related to the
inequality relations verified by the strength of interaction among
different components. Relevance of these results in bulk and neck
fragmentation is discussed.}

\section{Instabilities and fluctuations in ANM}

A binary system, including ANM, manifest a richer thermodynamical
behaviour as a consequence of the new degree of freedom. 
The phase transitions are more complex because they have
to accommodate one more conservation law. Also a variety of
thermal fluctuations with different components composition
can develop in the system.  

In particular, in symmetric nuclear matter (SNM) one encounters
two kinds of density fluctuations:
(i) isoscalar, when the densities
of the two components oscillate in phase with equal amplitude;
(ii)  isovector when the two densities fluctuate still with equal
amplitude but out of phase.
These are the normal modes of the system in the sense that
the variation of free energy density up to second order reduces to a normal
quadratic form when is expressed in terms of them. 
Mechanical instability is associated with
instability against isoscalar fluctuations leading to cluster
formation while chemical
instability is related to instability against isovector fluctuations,
leading to species separation.

A general analysis, for ANM, can be performed in the framework of the
Fermi liquid theory \cite{baym78}. We limit ourselves to monopolar
deformation for proton and neutron Fermi seas $\nu_p^{0}, \nu_n^{0}$
and consider here momentum independent interactions such that
$F^{q_1q_2}_{l=0}$ ($q_{i}=n,p$) is the only non-zero Landau parameter
Then the thermodynamical stability condition at $T=0$ is obtained as:
\begin{equation}
\delta H - \mu_p \delta \rho_p - \mu_n \delta \rho_n = \frac{1}{2}
(a {\nu_p^{0}}^{2} + b {\nu_n^{0}}^{2} + c \nu_p^{0} \nu_n^{0}) > 0~~
\label{varia1}
\end{equation}
where  $H$ is the energy density and
\begin{eqnarray}
a = N_p(0)(1 + F_{0}^{pp})~~;~
b = N_n(0)(1 + F_{0}^{nn})~~; \nonumber \\
c = N_n(0) F_{0}^{pn} + N_p(0) F_{0}^{np}.~~~~~~~~~~  \label{abc}
\end{eqnarray}
with  $N_{q}$ the
single-particle level density at the Fermi energy
We diagonalize the form of Eq.(\ref{varia1}) by introducing the following
transformation:
\begin{eqnarray}
u = cos\beta~ \nu_p^{0} + sin\beta~ \nu_n^{0},    \nonumber \\
v = - sin\beta~ \nu_p^{0} + cos\beta~ \nu_n^{0},         \label{rot}
\end{eqnarray}
where a {\it mixing} angle $0 \le \beta \le \pi/2$ is defined by
\begin{equation}
tg~ 2\beta = \frac{c}{a-b} = \frac{N_n(0) F_{0}^{pn} + N_p(0) F_{0}^{np}}
{N_p(0)(1 + F_{0}^{pp}) - N_n(0)(1 + F_{0}^{nn})}. \label{beta}
\end{equation}
Then Eq.(\ref{varia1}) takes the form
\begin{equation}
\delta H - \mu_p \delta \rho_p - \mu_n \delta \rho_n =
X u^2 + Y v^{2} \label{varia2}
\end{equation}
where
\begin{equation}
X = \frac{1}{2} (~a + b + sign(c) \sqrt{(a-b)^{2} + c^{2}}~) 
\equiv \frac{N_p(0)+N_n(0)}{2} (~ 1 + F_{0g}^s~) \label{A}
\end{equation}
and similar
\begin{equation}
Y= \frac{1}{2} (~a + b - sign(c) \sqrt{(a-b)^{2} + c^{2}}~) 
\equiv \frac{N_p(0)+N_n(0)}{2} (~ 1 + F_{0g}^a ~). \label{B}
\end{equation}

Thus with the rotation (\ref{rot}) we separate the total variation
Eq.(\ref{varia1}) into two independent contributions, the new "normal" modes,
characterized by a "mixing angle" $\beta$, which depends on the density of
states and  the interaction.
since $cos\beta, sin\beta$ are both positive, we interpret
$u$- and $v$-variations as new independent
${\it isoscalar}$-like and ${\it isovector}$-like fluctuations appropriate
for asymmetric systems. The proton and neutron densities will
fluctuate in phase for  isoscalar-like variations and out of
phase for isovector-like variations, see Eq.(\ref{rot}) and they reduces
to the usual isoscalar and isovector modes for SNM. 

From Eq.(\ref{varia2}) we see that thermodynamical stability requires
$X>0$ {\it and} $Y>0$. Equivalently, the following conditions have
to be fulfilled:
\begin{equation}
1 + F_{0g}^s > 0~~~~ and~~~~1 + F_{0g}^a > 0,          \label{pomegen}
\end{equation}
They represent Pomeranchuk stability conditions extended
to asymmetric binary systems.

\begin{figure}[htb]
\begin{minipage}[t]{55mm}
\epsfysize=5.cm
\centerline{\epsfbox{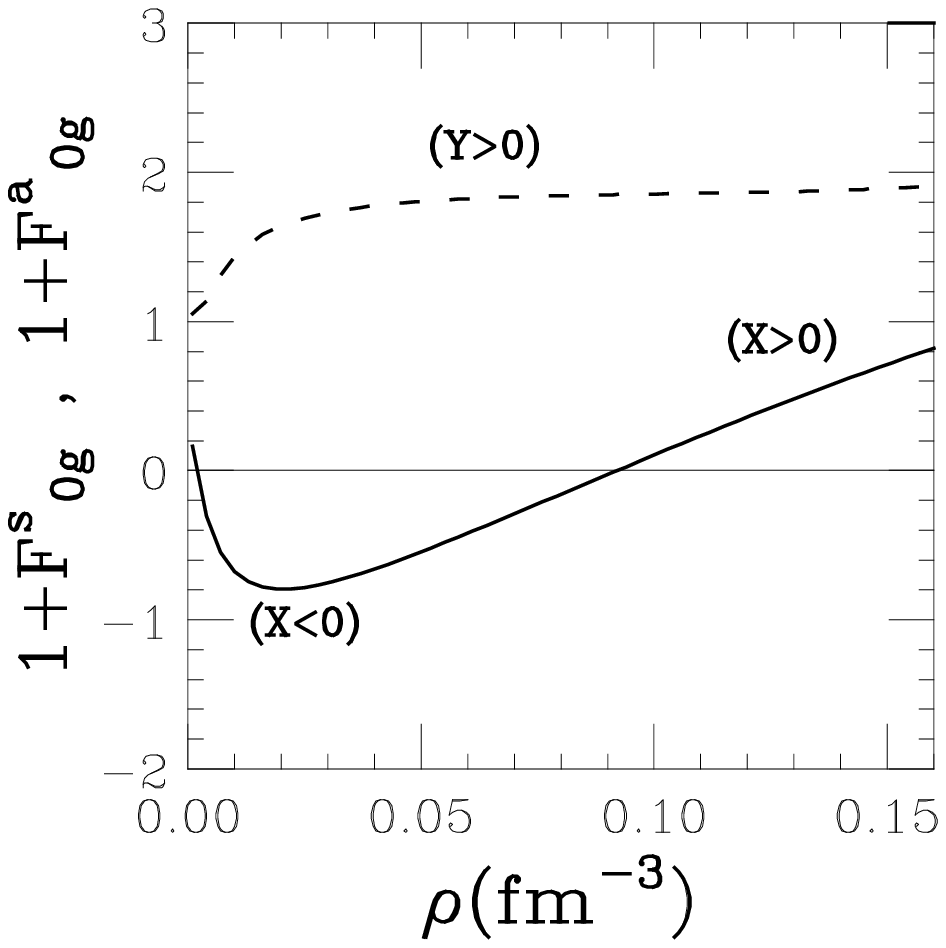}}
\caption{Density dependence of generalized Landau parameters for y=0.1
\label{fig_xy}}
\end{minipage}
\hspace{\fill}
\begin{minipage}[t]{55mm}
\epsfysize=4.65cm
\centerline{\epsfbox{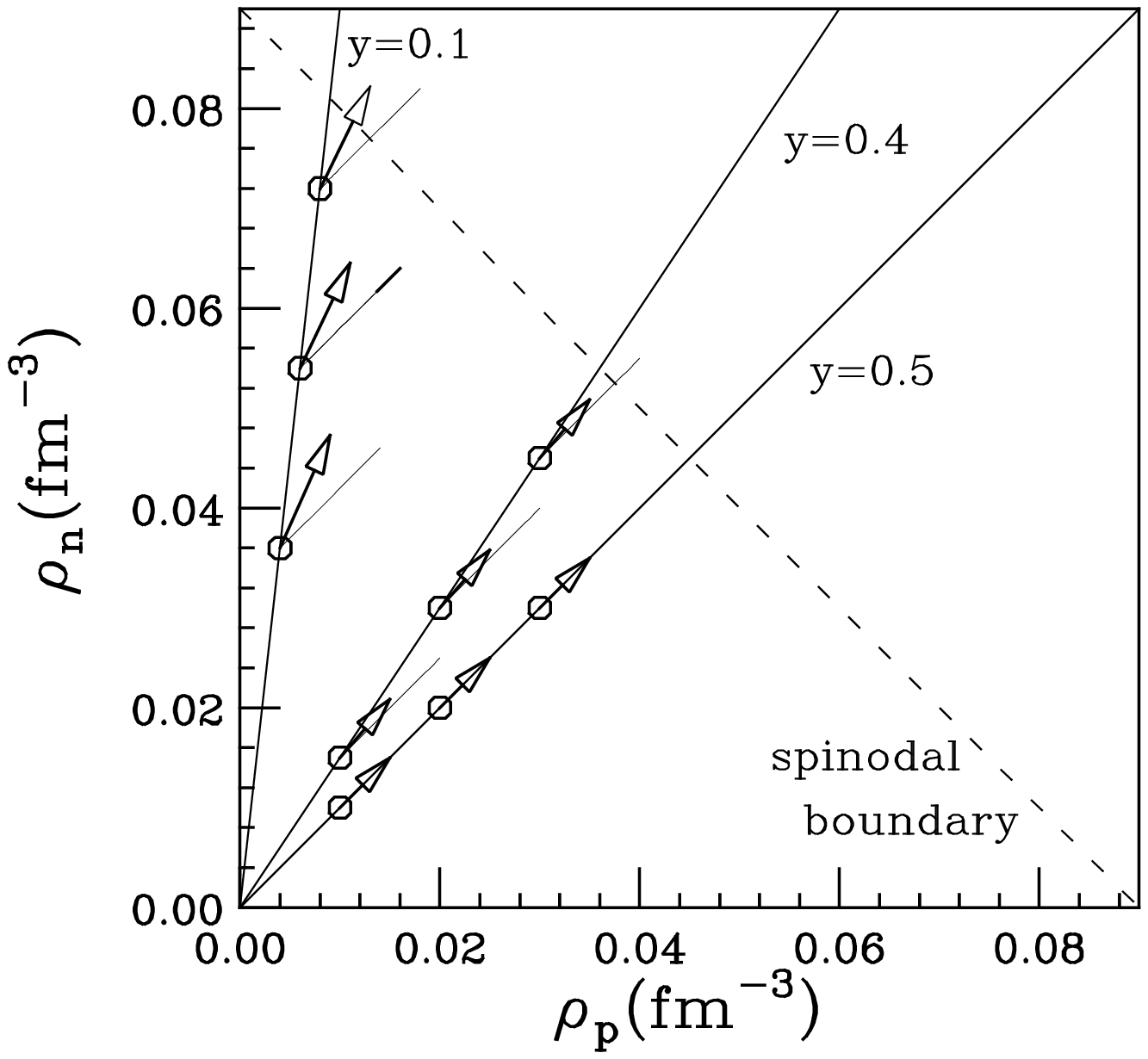}}
\caption{The unstable normal modes for liquid phase in the spinodal region.
\label{fig_ar}}
\end{minipage}
\end{figure}

The new stability conditions, Eq.(\ref{pomegen}),
are equivalent to mechanical and chemical stability of a
thermodynamical state, \cite{barranco80},
\cite{muller95}, \cite{baoan97}, i.e.
\begin{equation}
\left({\partial P \over \partial \rho}\right)_{T,y} > 0~~~and~~~
\left({\partial\mu_p \over \partial y}\right)_{T,P} > 0
\end{equation}
where $P$ is the pressure and $y$ the proton fraction, see \cite{baran01} .
Such general analysis leads to the conclusion
that in the low energy region, where $c<0$ as predicted by
all effective interactions, the asymmetric
nuclear matter is unstable only against isoscalarlike modes.
This is shown for a Skyrme-like interaction \cite{baran01} in Figure 
\ref{fig_xy},
but this effect is expected to by very robust, present for all 
interactions.

The direction of unstable mode in the plane $\rho_n-\rho_p$ in the liquid
phase is indicated by the arrows in the Figure~\ref{fig_ar}. The arrow 
picks toward less asymmetric liquid phase: the isospin distillation
take place. In this figure the length of the arrows
was drawn to correspond to the same proton density perturbation.
Therefore it indicate the "response" of neutrons and their
relative sizes allows a comparison of the effect 
throughout spinodal region. In the gas phase the asymmetry is larger
then the initial value, the arrows being oriented in opposite directions.
The tilting angle is a measure of isospin distillation which will depend on
the slope of the symmetry energy in the low energy region. Therefore the isospin
content of the fragments will be an interesting observable to constrain
the isovector part of the EOS, see the next sections.

\begin{figure}[htb]
\epsfysize=3.8cm
\centerline{\epsfbox{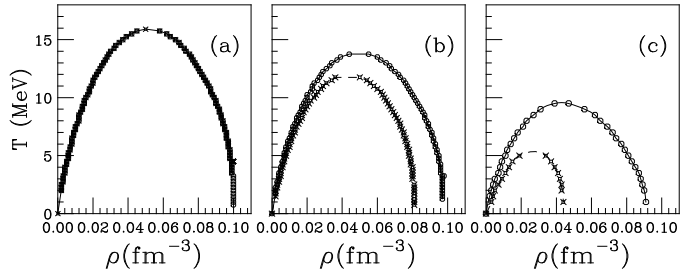}}
\caption{
Spinodal boundary of asymmetric nuclear matter(open circles) and
mechanical instability boundary (crosses) for three proton
fractions:(a) $y=0.5$,(a) $y=0.25$,(a) $y=0.1$
\label{fig_sp}} 
\end{figure}

The physical meaning of
the thermodynamical chemical and mechanical instabilities is connected to the
inequality relations existing between the interactions
among different species \cite{baran01}. Indeed the system 
will show up as a chemical instability if
$(-t a -b/t ) < c < -2\sqrt{ab}$ and as a
mechanical instability if $c < (-t a -b/t ) < -2\sqrt{ab}$, where
 $t = \frac{y}{1-y} \frac{N_n(0)}{N_p(0)}$.
The regions were each of these conditions are verified are indicated in 
Figure~\ref{fig_sp} as chemical and mechanical instability respectively.

\section{Kinetic of Phase Transition in ANM}

Contrary to the situation regarding the phase transition in macroscopic
systems were the observation time scale is much larger than the
time scale of microscopic process that leads to drops (bubbles) formation
(except for the critical point where are required days because of
slowing dawn phenomenon) for the heavy ions collision the reaction time
can be comparable to the fragment formation time. The violent collision
may quench the system in the instability region of phase diagram. The kinetic
mechanism responsible for the phase transition is the spinodal decomposition.
The  unstable fluctuations are amplified and growth exponentially
in the first stages leading to higher density domains. In asymmetric
nuclear matter they are especially interesting because of isoscalarlike
character of the unstable mode and presence of isospin distillation.

\begin{figure}[htb]
\epsfysize=4.4cm
\centerline{\epsfbox{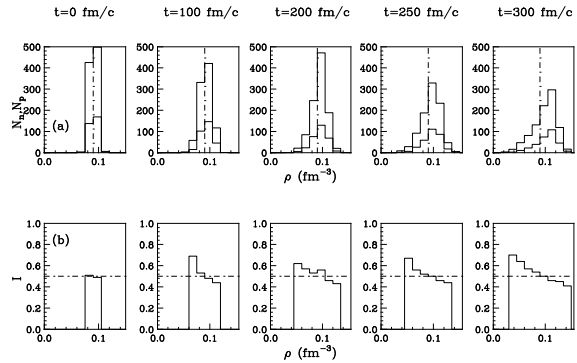}}
\caption{
Time dependence of the proton and neutron numbers (a)
and asymmetry (b) in different density bins.
\label{fig_as}} 
\end{figure}

We have studied the spinodal decomposition for ANM
in a box with side $L=~24fm$ for several  values of
the initial asymmetry and initial density
at temperature $T=5$ MeV. Several kinds of
initial 
perturbation were created automatically due to the random choice of
particle positions.
Results for the initial asymmetry 
$I=0.5$ and initial conditions corresponding to chemical
instability region ($\rho_{init}=0.09 fm^{-3}$)
are reported in Figure~\ref{fig_as} (a) and (b) respectively.
It is shown the time evolution of
neutron (thick histogram) and proton (thin histogram)
abundance Figure~\ref{fig_as} (a) and asymmetry ,Figure~\ref{fig_as} (b),
in various density bins.
The dashed line respectively shows the initial uniform density
value $\rho \simeq 0.6\rho_0$  and the initial asymmetry
$I=0.5$. The in-phase driving to higher density for
neutrons and protons is evident indicating the isoscalarlike character
of the instability. The consequence is the fragment formation. However the
clustering mechanism is accompanied by isospin distillation
leading to very different asymmetries, lower (higher) in the
liquid (gas) phases.

We want to stress that an identical qualitative behaviour
is observed if we start from the mechanical instability region
\cite{baran98}, in agreement
to the discussion of previous section.
 
\section{Isospin dynamics from bulk to neck fragmentation}
 
We follow the reaction dynamics for the collision of the system
$^{124}Sn+^{124}Sn$ at 50AMeV for two impact parameters $b=2fm$ and $b=6fm$
respectively. We consider an asystiff EOS corresponding to a
linear dependence with the density of the mean-field symmetry term.  

For
a semicentral collision, $b=2fm$, the reaction mechanism corresponds to 
bulk fragmentation. After a first compression phase (until about 40-50fm/c)
a fast expansion phase follows (until 110-120fm/c). Then during the
fragmentation stage the system will break up and the fragment formation 
process take place up to the freeze-out time (around 260-280fm/c). These
three main stages are characterized by specific features of the isospin
dynamics since the system explores different density regions.

\begin{figure}[htb]
\epsfysize=5.7cm
\centerline{\epsfbox{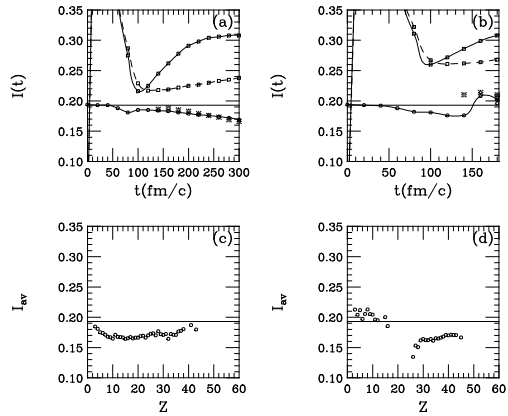}}
\caption{
Time dependence of the asymmetry in gas and liquid phases for b=2fm (a)
and b=6fm (b) as well as
the asymmetry of the primary fragments at freeze-out time for b=2fm (c)
and b=6fm (d).
\label{fig_iso1}} 
\end{figure}

In particular we focused on the time evolution of the asymmetry
of the "central" gas (solid line and squares), and total gas (dashed+squares),
as well as of the "central" liquid (solid+circles) and IMF
(clusters with $3<Z<23$ , stars), \ref{fig_iso1} (a).  
By "central" we mean a region having a linear dimension of the order
of $20fm$ corresponding to the active volume in which the fragmentation
take place.  

The spinodal mechanism is well evidenced by the inversion in the trend
for both liquid and gas phase corresponding to isospin distillation. The
isospin content of primary fragments is below the initial value
but also clearly lower than the value at the beginning of the phase transition,
 due to isospin distillation, see Figure~\ref{fig_iso1} (a) and (c).

 For b=6fm we observe a quite different behaviour. Now in the overlap
region a neck structure is developing. During the interaction time 
(from about 80-120fm/c) it heats and expands but remains in contact with
the denser and colder region of projectile-like (PLF) and target-like (TLF).
Now the surface instabilities of a cylindrical shape neck region and
the fast leading motion of the PLF and TLF will play an important role
in the mechanism of IMF production. A correspondingly different dynamics
of the isospin is evident from the Figure~\ref{fig_iso1} (b).
A larger asymmetry of IMF, Figure~\ref{fig_iso1} (d),
suggest a neck region richer (poorer) in neutrons (protons). This can be
related to proton and neutron migrations between more diluted neck region and
denser PLF and TLF as dictated by the chemical potential gradients of the
two species.

\section{Conclusion}

In the previous section in an approach based on a Fermi liquid theory
for two components systems we analyze several properties of ANM in low
density region of phase diagram. It is concluded that:

-in the low density region, of interest for nuclear liquid-gas phase
transition, the system can be characterized by a {\bf unique spinodal region},
see also \cite{baran01}, defined by the instability against isoscalarlike
fluctuations; inside this we can identify a mechanical instability
region and a chemical one;

-the physical meaning of chemical and mechanical instabilities is connected
to the inequality relations which are satisfied by  interactions
among different components; when this inequality changes the sign the chemical
instability transform in mechanical one;

-during the time development of the spinodal decomposition in ANM 
 the fragment formation accompanied by the isospin distillation
is observed.
The qualitative behaviour is unchanged when we move the initial conditions
from chemical instability region to mechanical instability one;

We have also made a connection of these features with isospin transport
properties in microscopic simulations of fragmentation reactions of charge
asymmetric ions in the medium energy range.
We observed a different isospin dynamics when we pass from
bulk fragmentation, for semi-central
collisions to neck fragmentation, in semi-peripheral collision.

We also mention that a dependence on the symmetry term of EOS was
evidenced for the observable showed above as well as for some other
as is discussed in more detail in the refs. \cite{baran02a},
\cite{colonna98}, \cite{scalone99}, \cite{ditoro02a},
\cite{drago02}.

\end{document}